\documentstyle[twoside,fleqn,espcrc2,psfig]{article}

\title{Perturbative coefficients for improved actions
       by Monte Carlo at large $\beta$}
\author{Howard D. Trottier\address{Physics Department, 
Simon Fraser University, Burnaby, B.C., CANADA V5A 1S6}
and G. Peter Lepage\address{Newman Laboratory of Nuclear Studies,
Cornell University, Ithaca, NY, 14853}}

\begin{document}

\begin{abstract}
Perturbative estimates of operator coefficients for improved
lattice actions are becoming increasingly important for precision
simulations of many hadronic observables. 
Following previous work by Dimm, Lepage, and Mackenzie, we consider 
the feasibility of computing operator coefficients from numerical 
simulations deep in the perturbative region of lattice theories. 
Here we introduce a background field technique that may allow
for the computation of the coefficients of clover-field operators in
a variety of theories. This method is tested by calculations of the 
renormalized quark mass in lattice NRQCD, and of the $O(\alpha_s)$ 
clover coefficient for Sheikholeslami-Wohlert fermions. First results 
for the coefficient of the magnetic moment operator in NRQCD are 
also presented.
\end{abstract}

\maketitle

\section*{}
Recent simulations of many hadronic observables using tadpole-improved 
actions, on both coarse and fine lattices, have yielded promising 
results. This has spurred further development of ever more highly 
improved lattice actions. At the same time, it has also become apparent 
that one must go beyond tree-level improvement in order to obtain 
precision results for many quantities. 

The increasing diversity and complexity of improved actions
makes traditional perturbative calculations of operator coefficients
in these actions problematic. A simpler alternative is needed,
if only to provide reasonable estimates of these coefficients. 
Dimm, Lepage, and Mackenzie have shown that the 
renormalized mass for Wilson fermions, and the energy zero in NRQCD, 
can be calculated from numerical simulations, deep in the 
perturbative region of these theories \cite{DLM}.

In this work we introduce a background field technique for
perturbative matching of lattice theories by numerical simulation.
We use this technique to estimate the coefficients of 
clover-field operators for both Sheikholeslami-Wohlert \cite{SW}
and NRQCD \cite{NRQCDUps} fermions.

The clover coefficients are tuned by computing quark propagators in 
a uniform background magnetic field, and matching the spin-flip 
energy
\begin{equation}
   \Delta E(gB) \equiv E(\downarrow,gB) - E(\uparrow,gB)
\end{equation}
in the lattice theory with the same quantity in continuum QCD. 
The continuum calculation has been done by Sapirstein \cite{Sap},
using the background field formalism \cite{Abbott}.
We used a lattice formulation of the background field
method. This method has the important feature that the 
combination $gB$ of bare coupling 
$g$ and bare field strength $B$ does not get renormalized. 
Thus lattice and continuum results for $\Delta E$ can
be directly compared, without regard to
the different regulators that are used.

We used the Wilson gluon action, and split the link variable 
into a classical background part and a fluctuating dynamical part,
$U_{{\rm tot},\mu}(x) = U_{{\rm dyn},\mu}(x) U_{{\rm cl},\mu}(x)$.
For an Abelian background (as used in the continuum \cite{Sap}),
$U_{{\rm cl},\mu=2}(x_1) = e^{i gB x_1 \lambda_3}$. Periodic
boundary conditions also require
$U_{{\rm cl},\mu=1}(x_1=N,x_2) = e^{-i gB N x_2 \lambda_3}$,
and $g B=2\pi n_{\rm ext}/N^2$, where $n_{\rm ext}$ is an integer.

This background field is an exact solution of the 
equations of motion for the effective action, with 
{\it zero\/} external current. This is because the only covariant 
vector operators that can be formed from the classical field strength 
$F_{{\rm cl},\mu\nu}$ are all identically zero 
(e.g., $D_\mu[A_{\rm cl}] F_{{\rm cl},\mu\nu} = 0$).
Consequently we do not need to make a 
perturbative calculation of the external current required to
support a more general classical field. 

On the other hand, a given background field does not minimize the 
effective action. In a numerical simulation one observes
``tunneling'' of the system, from the extremum at $F_{\rm cl}$,
to the minimum at zero field. To compute observables in the
presence of the classical field we modify the path integral,
introducing a ``current'' $J$ into the action as an intermediate step, 
in order to create a minimum in the effective potential near 
$F_{\rm cl}$:
\begin{equation}
   Z[J] \equiv \int \left[ d U_{\rm dyn} \right]
   e^{-\beta( S_{\rm Wil}[U_{\rm dyn} U_{\rm cl}]
             + \lambda_J J[U_{\rm dyn}] )} .
\label{ZJ}
\end{equation}
Physical quantities in the presence of the classical field
are obtained by extrapolation in the coupling, 
$\lambda_J \to 0$. However, one must 
perform this extrapolation from sufficiently large $\lambda_J$, in 
order for the system to be dominated by fluctuations 
in the dynamical fields near the classical field. 

\begin{figure}[htb]
\vspace{-9pt}
\psfig{file=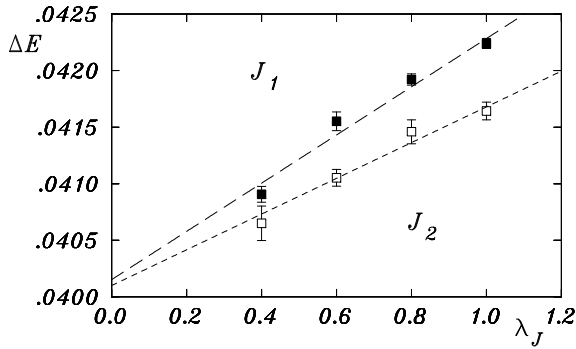,height=4.5cm,width=7.5cm,angle=90}
\vspace{-18pt}
\caption{NRQCD spin-flip energy for two external 
currents ($\beta=16$, $M_0=1.8$).}
\vspace{-15pt}
\label{fig:J1J2}
\end{figure}

To test this procedure, the NRQCD spin flip energy was computed 
twice, using two currents,
$J_1 = \sum_{x,\mu} \mbox{Re-Tr} 
       [ U_{{\rm dyn},\mu}(x) ]$,
and
$J_2 = (1/gB) \sum_{x,\mu} \mbox{Im-Tr}
       [ U_{{\rm dyn},\mu} F_{{\rm cl},12} ]$.
The extrapolated results agree, as shown in Fig. \ref{fig:J1J2}.

The action (and hence the current $J$) must be invariant 
under background field transformations, 
\begin{eqnarray}
\lefteqn{ 
   U_{{\rm cl},\mu}(x) \rightarrow
   \Omega(x) U_{\rm cl} \Omega^\dagger(x + \hat\mu) , } 
\nonumber \\
\lefteqn{ 
   U_{{\rm dyn},\mu}(x) \rightarrow
   \Omega(x) U_{\rm dyn}\Omega^\dagger(x) , }
\label{Gback}
\end{eqnarray}
in order to preserve the nonrenormalization of $gB$. 
Likewise, we require a background field-covariant gauge fixing 
prescription for the dynamical fields, when computing quark propagators:
\begin{eqnarray}
\lefteqn{
   \langle 0 \vert T\left( \psi(x) \bar\psi(0) \right) \vert 0 \rangle = }
\nonumber \\
\lefteqn{
   {1\over Z(J)} \int \left[ d U_{\rm dyn} \right] 
   e^{-\beta( S_{\rm W} + \lambda_J J )} 
   K^{-1} \!\! \left[ U_{\rm dyn}^G U_{\rm cl}; x \right] , }
\label{prop}
\end{eqnarray}
where $U_{\rm dyn}^G$ is the gauge-fixed dynamical field. We gauge 
fix with respect to transformations of the dynamical field 
(with fixed classical field)
\begin{equation}
   U_{{\rm dyn},\mu}(x) \rightarrow 
   \Omega(x) U_{\rm dyn} U_{\rm cl}(x) 
   \Omega^\dagger(x+\hat\mu) U_{\rm cl}^\dagger(x) .
\label{Gdyn}
\end{equation}
We used a background field axial ($U_{{\rm dyn},4} = I$) plus 
Coulomb ($\sum_{i=1}^3 D[U_{\rm cl}]_i U_{{\rm dyn},i} = 0$)
gauge fixing \cite{DLM}. These lattice transformations and 
gauge fixings have the desired continuum limits \cite{Abbott} 
(note that the quantum symmetry, Eq. (\ref{Gdyn}), is broken by 
the external current just as in the continuum).

\begin{figure}[htb]
\vspace{-9pt}
\psfig{file=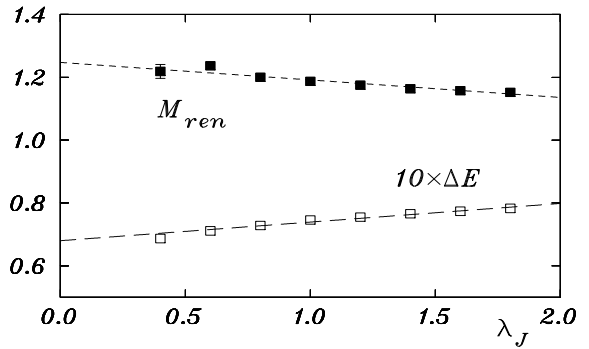,height=4.5cm,width=7.5cm,angle=90}
\vspace{-18pt}
\caption{NRQCD renormalized quark mass and spin-flip energy 
vs. $\lambda_J$ ($\beta=9$, $M_0=1.18$, with tadpole improvement).}
\vspace{-15pt}
\label{fig:NRQCD}
\end{figure}

We used this approach to calculate the renormalized
quark mass $M_{\rm ren}$ in lattice NRQCD. Relativistic corrections 
up to $O(v^4)$ in the mean quark velocity \cite{NRQCDUps}
were included. The renormalized masses were extracted from 
quark correlation functions with small nonzero three-momentum.
Sample results are shown in Fig. \ref{fig:NRQCD}. 
All simulations were done on $12^3 \times 24$ lattices 
at $\beta=9$, with $n_{\rm ext}=2$, and current $J_1$ was used. 
Simulations were done both with and without tadpole improvement
(using the average plaquette to compute $u_0$ in the former case).

We did linear extrapolations to $\lambda_J = 0$ to obtain the 
``true'' masses, given in Tables \ref{table:NRQCD} and
\ref{table:NRQCDu0}. 
Errors include a systematic error due to the extrapolation.
The Monte Carlo (``MC'') estimates of the masses are in
good agreement with results from perturbation theory (``PT''), 
due to Morningstar \cite{Colin}. Notice that most of
the renormalization is due to tadpoles, which is clearly
resolved by the Monte Carlo results.

\begin{table}[hbt]
\vspace{-6pt}
\setlength{\tabcolsep}{1.25pc}
\newlength{\digitwidth} \settowidth{\digitwidth}{\rm 0}
\catcode`?=\active \def?{\kern\digitwidth}
\caption{Renormalized quark mass and clover
coefficient in lattice NRQCD for three bare masses $M_0$,
{\it without\/} tadpole improvement.}
\label{table:NRQCD}
\begin{center}
\begin{tabular}{cccc}
\hline
  \\[-10pt]
  $M_0$  &  $M_{\rm ren}$  &  $M_{\rm ren}$  & $c_4^{(1)}$ \\
         &       (MC)      &     (PT)        &    (MC)     \\
\hline
  2.65   &      2.90(3)    &     2.94        &   4.8(2)    \\
  1.90   &      2.16(3)    &     2.20        &   4.0(2)    \\
  1.18   &      1.52(3)    &     1.51        &   2.3(2)    \\
\hline
\end{tabular}
\end{center}
\vspace{-18pt}
\end{table}

\begin{table}[hbt]
\vspace{-18pt}
\setlength{\tabcolsep}{1.25pc}
\caption{NRQCD renormalizations {\it with\/} tadpole improvement.}
\label{table:NRQCDu0}
\begin{center}
\begin{tabular}{cccc}
\hline
  \\[-10pt]
  $M_0$  &  $M_{\rm ren}$  &  $M_{\rm ren}$  & $c_4^{(1)}$ \\
         &       (MC)      &     (PT)        &    (MC)     \\
\hline
  2.65   &      2.69(3)    &     2.73        &   1.3(2)    \\
  1.90   &      1.95(3)    &     1.98        &   0.9(2)    \\
  1.18   &      1.24(3)    &     1.26        &   0.6(2)    \\
\hline
\end{tabular}
\end{center}
\vspace{-18pt}
\end{table}

Simulation results for $\Delta E$ in NRQCD are also shown in 
Fig. \ref{fig:NRQCD}. We find the coefficient of the magnetic moment 
operator, $c_4$, by matching the (extrapolated) lattice $\Delta E$
to the continuum value \cite{Sap}.
Defining $c_4 = 1 + c_4^{(1)} \alpha_s$, and using 
$\alpha_s=\alpha_V(\pi/a)$, we obtain the results in 
Tables \ref{table:NRQCD} and \ref{table:NRQCDu0}.

To calibrate our NRQCD calculations, we analyzed
spin-flip energies for Sheikholeslami-Wohlert (SW) fermions, 
where the clover coefficient $c_{sw}$ is known \cite{LW}.
We used a tree-level clover coefficient, and 
the value determined nonperturbatively for massless fermions
in Ref. \cite{LW} ($c_{sw}=1.24$ at $\beta=9$). We worked at 
modest quark masses ($ma \approx 0.5$) at which $O((ma)^2)$ errors 
may dominate over $O(\alpha_s ma)$ errors. This prevents a direct 
matching of the lattice results with the continuum. Instead, we compare 
the simulation results for $\Delta E$ with the spin-flip energy for 
``classical'' SW fermions [$U_{\rm dyn}=I$, $c_{sw}=1$,
and with the classical mass identified with 
$E(\uparrow,gB,\lambda_J)$ measured in the simulation]. 
We expect the effects of $O(a^2)$ errors will roughly cancel 
in this approximate matching procedure.

\begin{figure}[htb]
\vspace{-9pt}
\psfig{file=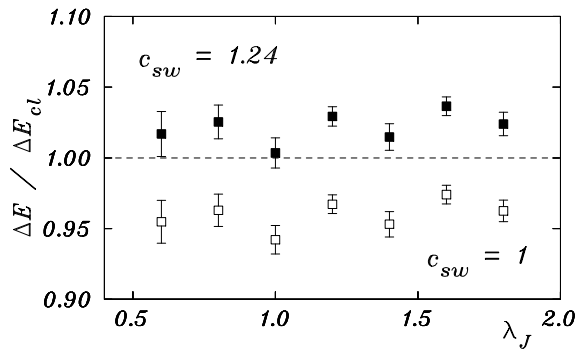,height=4.5cm,width=7.5cm,angle=90}
\vspace{-18pt}
\caption{Monte Carlo $\Delta E$ for SW fermions, compared to the 
``classical'' spin-flip energy $\Delta E_{\rm cl}$ 
($\beta=9$, $\kappa=.115$).}
\vspace{-15pt}
\label{fig:SW}
\end{figure}

Sample results are shown in Fig. \ref{fig:SW}. The simulation
results with a tree-level clover coefficient show a significant
departure from the matching condition. At this quark
mass ($ma \approx .5$), a value of $c_{sw}$ somewhat smaller than
the coefficient for massless fermions \cite{LW} is suggested
by these results.

In summary, we have introduced a background field method for simulations 
in the perturbative region of lattice theories. Encouraging preliminary
results were obtained for the coefficients of clover operators for 
SW and NRQCD fermions, and for the renormalized quark mass in NRQCD.

\end{document}